\documentclass[10pt,letterpaper,twocolumn]{article} 
\usepackage{ol2}  
\usepackage{amsmath}
\usepackage{bm}
\usepackage{amscd}
\usepackage{amssymb}
\usepackage{graphicx}
\newcommand{\di}{\mathrm{d}}

\newcommand{\bbf}{\bm{f}}
\newcommand{\bx}{\bm{x}}
\newcommand{\by}{\bm{y}}
\newcommand{\bz}{\bm{z}}
\newcommand{\bk}{\bm{k}}

\newcommand{\bR}{\bm{R}}
\newcommand{\bK}{\bm{K}}

\newcommand{\bA}{\bm{A}}

\providecommand{\abs}[1]{\lvert#1\rvert}

\begin{document}
\twocolumn[ 
\title{Goos-H\"{a}nchen and Imbert-Fedorov shifts of a nondiffracting Bessel beam 
}
%
%
%
\author{Andrea Aiello,$^{1,2*}$  J. P. Woerdman,$^{3}$}
\address{$^1$ Max Planck Institute for the Science of Light, G\"{u}nter-Scharowsky-Str. 1/Bau 24, 91058 Erlangen, Germany}
\address{$^2$Institute for Optics, Information and Photonics, University Erlangen-N\"{u}rnberg,\\ Staudtstr. 7/B2, 91058 Erlangen, Germany.}
\address{$^3$Huygens Laboratory, Leiden University,
P.O.\ Box 9504, 2300 RA Leiden, The Netherlands}
\address{$^*$Corresponding author: andrea.aiello@mpl.mpg.de}
\begin{abstract}
Goos-H\"{a}nchen and Imbert-Fedorov shifts are diffractive corrections to geometrical optics that have been extensively studied for a Gaussian beam that is reflected or transmitted by a dielectric interface. Propagating in free space before and after reflection or transmission, such a Gaussian beam spreads due to diffraction. We address here the question how the Goos-H\"{a}nchen and Imbert-Fedorov shifts behave for a ``nondiffracting'' Bessel beam.
\end{abstract}



 ] 

\maketitle 
It has been known since a long time that the behavior of a finite-diameter light beam in  reflection and transmission at a dielectric interface differs from the predictions of geometrical optics. Due to diffractive corrections the beam is shifted in directions parallel and perpendicular to the plane of incidence \cite{JacksonBook}. The parallel shift is known as the Goos-H\"{a}nchen (GH) effect \cite{GH2} and the transverse shift as the Imbert-Fedorov (IF) effect \cite{Fedorov,Imbert}. These effects have been extensively studied \cite{Artmann,BliokhPRL,BliokhPRE}, not only for total internal reflection which is the context wherein the GH and IF effects were originally addressed, but also in partial dielectric reflection and transmission \cite{AielloOL08}. We note that the IF effect is closely related to the Spin Hall Effect of Light (SHEL) \cite{OnodaEtAlPRL}, \cite{BliokhPRL,HostenandKwiat,PhysRevLett.103.100401}. Further generalizations concern angular varieties of the GH and IF effects; these are observed in the far field of the reflected (or transmitted) beam \cite{NatPhoton.3.337}. Recently, also the influence of Orbital Angular Momentum of the incident beam on these diffractive shifts has been investigated \cite{Fedoseyev08,Bliokh:09,AielloOL2009}; for the transverse case this may be called the Orbital Hall Effect of Light (OHEL) \cite{AielloPra2010}.

The \emph{diffractive} origin of these effects raises the question how they behave when the incident beam is a so-called \emph{nondiffracting} Bessel beam. Such beams were perceived by Durnin \textit{et al.} as propagation-invariant solutions of the free-space scalar wave equation \cite{BB1,BB2}. These solutions have amplitudes proportional to Bessel functions. The zero-order Bessel beam has a bright central maximum (``needle beam'') which propagates in free space without diffractional spreading; the higher-order beams have a dark central core. Most work on Bessel beams is restricted to the paraxial limit \cite{BB3,BB4} but also the nonparaxial case (described by the Helmholtz wave equation) has been reported \cite{BB5,BB6}. 

Ideal Bessel beams have infinite transverse diameter and can therefore not be generated experimentally. However, there exist several experimental methods to generate finite-diameter approximations to a Bessel beam; these propagate over a finite axial distance in a nondiffracting manner (i.e. over distances much larger than the Raleigh length corresponding to the needle-beam diameter) \cite{BB7}. So, one may speculate that such a needle beam corresponds to a geometrical-optics ray that would \emph{not} show GH and IF shifts. Verification or rebuttal of this speculation requires proper theory; this is reported in the present paper.

Let us begin by briefly recalling to the reader what a Bessel beam is.
The \emph{scalar} $m^\text{th}$-order Bessel beam is a cylindrically symmetric monochromatic optical beam whose electric field has the following form:
\begin{align}\label{eq10}
E(R, \varphi,z) = & \; J_m(K_0 R)e^{i m \varphi}e^{i z \sqrt{k_0^2 - K_0^2} }\nonumber \\
\equiv & \; A(R, \varphi)e^{i z \sqrt{k_0^2 - K_0^2}},
 \end{align}
where $m$ is an integer number that fixes the value of the orbital angular momentum (OAM) of the beam, and $(R, \varphi,z)$ are the cylindrical spatial coordinates defined with respect to the main axis of propagation $\hat{{\bf z}}$:
\begin{align}\label{eq20}
\left\{\begin{array}{rcl}
         x & \! = & \! R  \cos \varphi, \\
         y & \! = & \! R \sin \varphi . 
       \end{array}
 \right.
 \end{align}
For a needle beam one has $m =0$.
In Eq. (\ref{eq10}) $k_0>0$ and $0 \leq K_0 \leq k_0 $ are two \emph{independent} parameters, where $K_0$  determines the \emph{angular width} $\vartheta_0$ of the central lobe (cone) of the
corresponding Bessel function via the definition 
\begin{align}\label{eq30}
K_0 = k_0 \sin \vartheta_0, \qquad  (0 \leq \vartheta_0 \leq \pi/2) .
 \end{align}
Throughout this Letter we will consider only \emph{paraxial} Bessel beams characterized by the condition 
\begin{align}\label{eq40}
\sin \vartheta_0 = {K_0}/{k_0} \ll 1.
 \end{align}
 The physical meaning of the angle $\vartheta_0$ is illustrated in Fig. 1 below.
%
%
\begin{figure}[h!]
\begin{center}
\includegraphics[angle=0,width=5truecm]{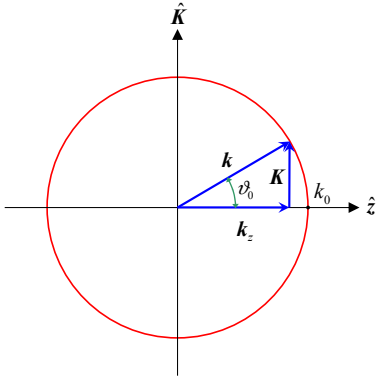}
\caption{\label{fig:1} Illustrating the physical meaning of the angle $\vartheta_0$. }
\end{center}
\end{figure}
%
%
It should be noticed that while $E(R, \varphi,z)$ is an exact solution to the Helmholtz equation
 $(\partial_x^2 + \partial_y^2 + \partial_z^2 + k_0^2) E =0$ in free space, the amplitude $A(R, \varphi)$ satisfies the
reduced equation  ${(\partial_x^2 + \partial_y^2  + K_0^2) A =0}$.

For actual calculations of both GH and IF shifts it is convenient to work in Fourier space and calculate the Fourier transform $\widetilde{A}(K, \phi)$ of the amplitude $A(R, \varphi) = J_m(K_0 R)e^{i m \varphi}$ as:
\begin{align}\label{eq70}
J_m(K_0 R)e^{i m \varphi} = \frac{1}{2 \pi} \int \widetilde{A}(k_x,k_y) e^{i \bK \cdot \bR} \, \di k_x \di k_y,
 \end{align}
where
\begin{align}\label{eq80}
\widetilde{A}(k_x,k_y) = \widetilde{A}(K, \phi) = \frac{1}{i^m K_0} \delta \left(K - K_0 \right) e^{i m \phi},
 \end{align}
with $K = (k_x^2 + k_y^2)^{1/2}$, $\bK \cdot \bR = x k_x + y k_y = K R \cos(\phi - \varphi)$,  and $k_x =K  \cos \phi, \; k_y = K \sin \phi$. It is worth noticing that in the literature Eq. (\ref{eq80}) is often written in spherical coordinates $(k_0,\vartheta,\phi)$ with $K = k_0 \sin \vartheta$, $K_0 \delta \left( K - K_0 \right) =  \delta \left( \vartheta - \vartheta_0 \right)/{\cos \vartheta_0 }$, and $\di k_x \di k_y \, = k_0^2 \sin \vartheta \cos \vartheta \, \di \vartheta \di \phi$.

Having written explicitly the Fourier representation of a scalar Bessel beam, we can now proceed as in \cite{AielloPra2010} and write the Fourier amplitude of a \emph{vector} Bessel beam as:
\begin{align}\label{eq85}
\widetilde{A}(k_x,k_y) \rightarrow   \widetilde{\bA}(k_x,k_y) = \bbf_{\perp}(\bK)\widetilde{A}(k_x,k_y),
\end{align}
where $\bbf_{\perp}(\bK) = \hat{\bbf} - \hat{\bk}(\hat{\bk} \cdot \hat{\bbf})$, with $\hat{\bk} = k_x \hat{\bx} + k_y \hat{\by} + (k_0^2 - K^2)^{1/2} \hat{\bz}$ and $\hat{\bbf} = f_p \hat{\bx} + f_s \hat{\by}, \; \abs{\hat{\bbf}}^2=1$. Here, according to  \cite{AielloPra2010}, the three unit vectors $\{ \hat{\bx},\hat{\by},\hat{\bz} \}$ form a right-handed Cartesian reference frame attached to the incident beam propagating along the axis $\hat{\bz}$.

At this point the GH and IF shifts for a Bessel beam impinging at the angle $\theta$ upon a planar interface may be straightforwardly calculated by using the formulas given in Sec. III of Ref. \cite{AielloGH2009}. Once again, following Ref. \cite{AielloPra2010} we define the ``intrinsic'' (namely, beam-independent) longitudinal and transverse shifts as, respectively,
\begin{align}\label{eq130}
X_ \lambda = - i \frac{\partial \ln r_ \lambda}{\partial \theta} = {\phi_ \lambda}' - i \frac{{R_ \lambda}'}{{R_ \lambda}},
 \end{align}
 and
\begin{align}\label{eq140}
Y_p =   i \frac{f_p}{f_s} \left(1 + \frac{r_s}{r_p} \right), \qquad
Y_s = & \;  -i \frac{f_s}{f_p} \left(1 + \frac{r_p}{r_s} \right),
 \end{align}
where $r_\lambda = R_\lambda \exp (i \phi_\lambda ), \; \lambda \in \{p,s\}$,  with the prime indicating derivatives with respect to the incidence angle $\theta$.
Moreover, we define the relative reflected energies $w_p$ and $w_s$ as:
\begin{align}\label{eq150}
w_p = \frac{ {a_p}^2  {R_p}^2  }{ {a_p}^2  {R_p}^2+ {a_s}^2  {R_s}^2}, \; w_s = \frac{ {a_s}^2  {R_s}^2  }{ {a_p}^2  {R_p}^2+ {a_s}^2  {R_s}^2},
 \end{align}
and the complex-valued longitudinal and transversal shifts $\Xi$ and $\Psi$ respectively, as:
\begin{align}\label{eq160}
\Xi = w_p X_p + w_s X_s, \qquad  \Psi = w_p Y_p + w_s Y_s,
 \end{align}
where
\begin{subequations}\label{Shift}
\begin{align}
\text{Re} \left( \Xi \right) = &  \;  \frac{ {a_p}^2  {R_p}^2  { \phi_p}'+ {a_s}^2  {R_s}^2  { \phi_s}'}{ {a_p}^2  {R_p}^2+ {a_s}^2  {R_s}^2} ,\label{ShiftA} \\
\text{Im} \left( \Xi \right) = &  \;
-\frac{ {a_p}^2  {R_p} {R_p}'+ {a_s}^2  {R_s}  {R_s}'}{ {a_p}^2 {R_p}^2+ {a_s}^2  {R_s}^2},
\label{ShiftB} \\
\text{Re} \left( \Psi \right) = &  \;
-\frac{ {a_p}  {a_s}  \cot \theta \left( {R_p}^2+ {R_s}^2\right) \sin \eta }{ {a_p}^2 {R_p}^2+ {a_s}^2  {R_s}^2} \nonumber \\
 &   \;
-\frac{ {a_p}  {a_s}  \cot \theta \left[2 {R_p}  {R_s}  \sin(\eta - { \phi_p}+ { \phi_s})\right]}{ {a_p}^2 {R_p}^2+ {a_s}^2  {R_s}^2},
\label{ShiftC} \\
\text{Im} \left( \Psi \right) = &  \;
\frac{ {a_p}  {a_s} \cot \theta \left( {R_p}^2- {R_s}^2\right)  \cos \eta }{ {a_p}^2 {R_p}^2+ {a_s}^2  {R_s}^2}. \label{ShiftD}
 \end{align}
 \end{subequations}

From Ref. \cite{AielloOL2008} it immediately follows that the ``traditional'' GH and IF shifts evaluated for a fundamental Gaussian beam of waist $w_0$ and Rayleigh range $L = k_0 w_0^2/2$ can be expressed in terms of the formulas given above as:
\begin{subequations}\label{eq170}
\begin{align}
k_0 \langle x_r \rangle   = & \;  \text{Re}(\Xi) + \frac{z_r}{L} \text{Im}(\Xi) ,\label{eq170A} \\
k_0 \langle y_r \rangle   = & \;  \text{Re}(\Psi) + \frac{z_r}{L} \text{Im}(\Psi) ,
\label{eq170B}
 \end{align}
 \end{subequations}
where the real parts of $\Xi$ and $\Psi$ give the spatial shifts of the beam, and the imaginary parts furnish the angular GH and IF shifts defined as $\partial  \langle x_r \rangle/ \partial z_r$ and $\partial  \langle y_r \rangle/ \partial z_r$, respectively. It should be noticed that here and in the subsequent formulas, according to  Ref. \cite{AielloPra2010},  the three Cartesian coordinates $\{ x_r, y_r, z_r\}$ are referred to a reference frame attached to the reflected beam of central wavevector $\tilde{\bk}_0$, with $z_r$ directed along $\tilde{\bk}_0$. A straightforward calculation shows that in the case of a  $m^\text{th}$-order Bessel beam, Eqs. (\ref{eq170}) become:
\begin{subequations}\label{eq180}
\begin{align}
\left. k_0 \langle x_r \rangle \right|_{z_r =0}  = & \;  \text{Re}(\Xi) - m \, \text{Im}(\Psi) ,\label{eq180A} \\
\left. k_0 \langle y_r \rangle \right|_{z_r =0}   = & \;  \text{Re}(\Psi) + m \, \text{Im}(\Xi) ,
\label{eq180B}
 \end{align}
 \end{subequations}
for the spatial part, and 
\begin{subequations}\label{eq190}
\begin{align}
 \frac{\partial \langle x_r \rangle}{\partial z_r}   = & \; \sin \vartheta_0^2 \, \text{Im}(\Xi),\label{eq190A} \\
 \frac{\partial \langle y_r \rangle}{\partial z_r}   = & \; \sin \vartheta_0^2 \, \text{Im}(\Psi),
\label{eq190B}
 \end{align}
 \end{subequations}
for the angular part. These formulas are the main result of this Letter. Before proceeding with the discussion of these formulas, a caveat is in order here. They were derived in straightforward manner by analogy with the Gaussian beam case. However, while a Gaussian beam is describable by means of normalizable functions, a Bessel beam does not. Luckily, in the practical calculation of GH and IF shifts, infinities present in the first-order moments of the electric field energy density distribution are exactly (and, unambiguously) compensated by the infinities given by the electric field energy density integrate over the whole space. Thus, the non-normalizable nature of (theoretical) Bessel beams does not represent a problem.

Some relevant issues follow from Eqs. (\ref{eq180}-\ref{eq190}) above.
First, since for a fundamental Gaussian beam of angular aperture $\theta_0$ one has $1/(k_0 L) = \theta_0^2/2$, then from Eqs. (\ref{eq190}) with $\sin \vartheta_0 \sim \vartheta_0$, it follows that the angular shift of a Bessel beam is about \emph{twice} the corresponding shift of a Gaussian beam. Second, Eq. (\ref{eq180}) shows a spatial/angular mixing analogous to the one present for a Laguerre-Gauss beam of OAM $m$ \cite{AielloPra2010}. Third, in particular for the case $m=0$,  the formulas derived above have profound consequences 
 for the (nondiffracting) core of the Bessel beam, i.e. the needle beam. The key point is that in first order perturbation theory (with respect to the expansion parameter $\vartheta_0$) the (full) Bessel beam is \emph{not} deformed upon reflection, so it translates rigidly as a whole. Since, as we saw above, the full Bessel beam shows the standard GH and IF shifts this must also be the case for its central nondiffracting core. The only escape from this conclusion is to leave the paraxial approximation and
 to go to second-order  (and higher)  perturbation orders (i.e.  relatively strong focusing); this reduces the length over which diffraction is effectively absent to a propagation length of the order of the diameter of the full Bessel beam. However, a treatment of beam shifts in this regime is outside the scope of this paper.

\bigskip
A.A. acknowledges support from the Alexander von Humboldt
Foundation.


\begin{thebibliography}{10}

\bibitem{JacksonBook}
J.~D. Jackson.
\newblock {\em Classical Electrodynamics}.
\newblock John Wiley {\&} Sons, New York, 3 edition, 1998.

\bibitem{GH2}
F.~Goos and H.~H\"{a}nchen.
\newblock Ein neuer and fundamentaler versuch zur total reflection.
\newblock {\em Ann. Phys. (Leipzig)}, 1:333, 1947.

\bibitem{Fedorov}
F.~I. Fedorov.
\newblock Theory of total reflection.
\newblock {\em Dokl. Akad. Nauk SSSR}, 105:465, 1955.

\bibitem{Imbert}
Christian Imbert.
\newblock Calculation and experimental proof of the transverse shift induced by
  total internal reflection of a circularly polarized light beam.
\newblock {\em Phys. Rev. D}, 5(4):787, 1972.

\bibitem{Artmann}
K.~Artmann.
\newblock Berechnung der seitenversetzung des totalreflektierten strahles.
\newblock {\em Ann. Phys. (Leipzig)}, 2:87, 1948.

\bibitem{BliokhPRL}
Konstantin~Yu. Bliokh and Yury~P. Bliokh.
\newblock Conservation of angular momentum, transverse shift, and spin Hall
  effect in reflection and refraction of an electromagnetic wave packet.
\newblock {\em Phys. Rev. Lett.}, 96:073903, 2006.

\bibitem{BliokhPRE}
K.~Yu. Bliokh and Yu.~P. Bliokh.
\newblock Polarization, transverse shifts, and angular momentum conservation
  laws in partial reflection and refraction of an electromagnetic wave packet.
\newblock {\em Phys. Rev. E}, 75:066609, 2007.

\bibitem{AielloOL08}
A.~Aiello and J.~P. Woerdman.
\newblock Role of beam propagation in Goos-H\"{a}nchen and Imbert-Fedorov
  shifts.
\newblock {\em Opt. Lett.}, 33(13):1437, July 2008.

\bibitem{OnodaEtAlPRL}
Masaru Onoda, Shuichi Murakami, and Naoto Nagaosa.
\newblock Hall effect of light.
\newblock {\em Phys. Rev. Lett.}, 93(8):083901, 2004.

\bibitem{HostenandKwiat}
Onur Hosten and Paul Kwiat.
\newblock Observation of the spin Hall effect of light via weak measurements.
\newblock {\em Science}, 319:787, 2008.

\bibitem{PhysRevLett.103.100401}
Andrea Aiello, Norbert Lindlein, Christoph Marquardt, and Gerd Leuchs.
\newblock Transverse angular momentum and geometric spin Hall effect of light.
\newblock {\em Phys. Rev. Lett.}, 103(10):100401, Aug 2009.

\bibitem{NatPhoton.3.337}
M.~Merano, A.~Aiello, M.~P. van Exter, and J.~P. Woerdman.
\newblock Observing angular deviations in the specular reflection of a light
  beam.
\newblock {\em Nat. Photon.}, 3:337, 2009.

\bibitem{Fedoseyev08}
V.~G. Fedoseyev.
\newblock Transformation of the orbital angular momentum at the reflection and
  transmission of a light beam on a plane interface.
\newblock {\em J. Phys. A}, 41(50):505202, 2008.

\bibitem{Bliokh:09}
Konstantin~Y. Bliokh, Ilya~V. Shadrivov, and Yuri~S. Kivshar.
\newblock Goos-H\"{a}nchen and Imbert-Fedorov shifts of polarized vortex
  beams.
\newblock {\em Opt. Lett.}, 34(3):389-391, 2009.

\bibitem{AielloOL2009}
A.~Aiello, M.~Merano, and J.~P. Woerdman.
\newblock Brewster cross polarization.
\newblock {\em Opt. Lett.}, 34:1207, 2009.

\bibitem{AielloPra2010}
M.~Merano, N.~Hermosa, J.~P. Woerdman, and A.~Aiello.
\newblock How orbital angular momentum affects beam shifts in optical
  reflection.
\newblock {\em Phys. Rev. A}, 82:023817, 2010.

\bibitem{BB1}
J.~E. Durnin.
\newblock Exact solutions for nondiffracting beams. I. The scalar theory.
\newblock {\em J. Opt. Soc. Am. A}, 4:651, 1987.

\bibitem{BB2}
J.~J.~Miceli J.~E.~Durnin and J.H. Eberly.
\newblock Diffraction-free beams.
\newblock {\em Phys. Rev. Lett.}, 54:1499, 1987.

\bibitem{BB3}
Y.~Z. Umul.
\newblock Apertured paraxial Bessel beams.
\newblock {\em J. Opt. Soc. Am. A}, 27:390, 2010.

\bibitem{BB4}
G.~S.~McDonald S.~Chavez-Cerda and G.~H.~C. New.
\newblock Nondiffracting beams: travelling, standing, rotating and spiral
  waves.
\newblock {\em Opt. Commun.}, 123:225, 1996.

\bibitem{BB5}
M.~Santarsiero R.~Borghi and M.~A. Porras.
\newblock Nonparaxial Bessel-Gauss beams.
\newblock {\em J. Opt. Soc. Am. A}, 18:1618, 2001.

\bibitem{BB6}
J.~Chen and Y.~Yu.
\newblock The focusing property of vector Bessel-Gauss beams by a high
  numerical aperture objective.
\newblock {\em Opt. Commun.}, 283:1655, 2010.

\bibitem{BB7}
J.~Arlt and K.~Dholakia.
\newblock Generation of high-order Bessel beams by use of an axicon.
\newblock {\em Opt. Commun.}, 177:297, 2000.

\bibitem{AielloGH2009}
A. Aiello, and J. P. Woerdman, Theory of angular Goos-H\"{a}nchen shift near
  Brewster incidence, {arXiv:0903.3730v2 [physics.optics]} (2009).

\bibitem{AielloOL2008}
Andrea Aiello and J.~P. Woerdman.
\newblock Role of beam propagation in Goos-H\"{a}nchen and Imbert-Fedorov
  shifts.
\newblock {\em Opt. Lett.}, 33(13):1437, July 2008.

\end{thebibliography}
\end{document}